# Single parameter model of marriage/divorces dynamics with countries classification

Andrzej Jarynowski[1,2], Piotr Nyczka[3,4]

*1. Smoluchowski Institute, Jagiellonian University, Cracow, Poland*
*2. Institute of Interdisciplinary Research, Wroclaw, Poland*
*3. UNESCO Chair of Interdisciplinary Studies, University of Wroclaw, Poland*
*Jabobs University, Bremen, Germany*

---

**Abstract**

We investigate a marriage/divorces problem via agent - based modeling. Current value of socio-economic pressure p (main model parameter) drives the dynamic of first marriage, remarriage or spontaneously marriage breaks up. Model reflects the behavior of the heterosexual population (frequency of changing partners, the ratio of singles in society). Theoretical agent-based simulation (with population approaches, e. g. births and deaths) are supplemented by historical values of divorces/marriages in various countries of the world from United Nation Registry and World Value Survey. In the model, agents have a list of their attributes and preference of a potential partner with Manhattan distance measure applied as a matching function. In the deterministic part of the model randomly selected agents could entry into a new relationship, or exchange partners, with respect to the matching distance and pressure parameter. Spouses could also spontaneously fall apart with a pressure depended probability. This simple model, although it assumes homogeneity of agents (on bipartite graph), and has only one key parameter, gives estimate of the socio-economic pressure values for different societies at given time point, knowing the frequency of changing partners, the percentage of married in society.

---

## 1. Introduction

In this study, we address the effect of socio-economic pressure on marriage building. We use agent-based modelling (ABM) to explore what is level of the pressure in worlds society under certain simplified conditions. We revisited problem well known in mathematics: the stable marriage problem [1], which is also widely used in other fields mostly in economics (Memorial Nobel Prize in 2012). In this problem, two sets of agents (e.g. men and women) must be matched pairwise in accordance to their mutual preferences. However, so far only little attention of the scientist community representing computational and natural science has been paid to construct tunable models to reproduce marriage-divorce properties observed in the real world. For example, there are algorithms for generation of the optimal and stable solution [2], but none (up to our knowledge) investigate dynamics of coupling/decoupling in societies from this perspective. On the other hand, most of sociological and psychological studies focus just on finding explanatory factors driving peoples decision to marry or divorce for given society, for given point in time [3]. We want to combine both methodologies to build simple model with one abstract parameter (pressure/freedom) describing chosen societies in general framework. It allow us to compare different societies and observe their properties (evolution in time). The main aim of this paper is to enhance the basic rules in marriage dynamics and predict trends in dynamically changing real societies.

### *1.1. Demographical and sociological background*

Research on the marriages and divorces has attracted a great interest, mostly due to importance on human wellbeing. Given intellectual heritage of registered-based and survey-based research social scientists build advanced theories [4]. We assume that macro-level concepts as social norms imply microlevel interactions as matching strategies. Decisions of individuals effect macro-level variables as divorce rates. Our pressure parameter plays a combined role social norms and economic benefits of staying in marriage. It can be understood as cultural transmission (macro-variable), while peer influence (micro-variable) is negligible. Despite there is only one tunable parameter in our model, the most important theoretical constrains are satisfied.
A) Second demographical transitions
The second demographic transitions refer (among others) to the changes in divorce numbers that occurred in Western civilization recently [5]. It succeeds first demographic transition characterized by strengthening marriage and strict divorce legislation. Since the second half of the twentieth century, the second transition in Western civilization witnessed the end of a long period of low divorce rates [6], and the principle of a unique, life-long legal partnership was questioned. Technically, the social regulatory mechanism change, gives a way of individual freedom of choice driven both by emotions and economic rationality. Transition as a socio-demographical process has a typical time scale of generation. As the first and the second transition are currently ongoing in various civilizations of the world, it's important to classify societies to processes stages. In this paper, we try to discuss, current situation in sampled countries by estimating local pressure/freedom parameters.

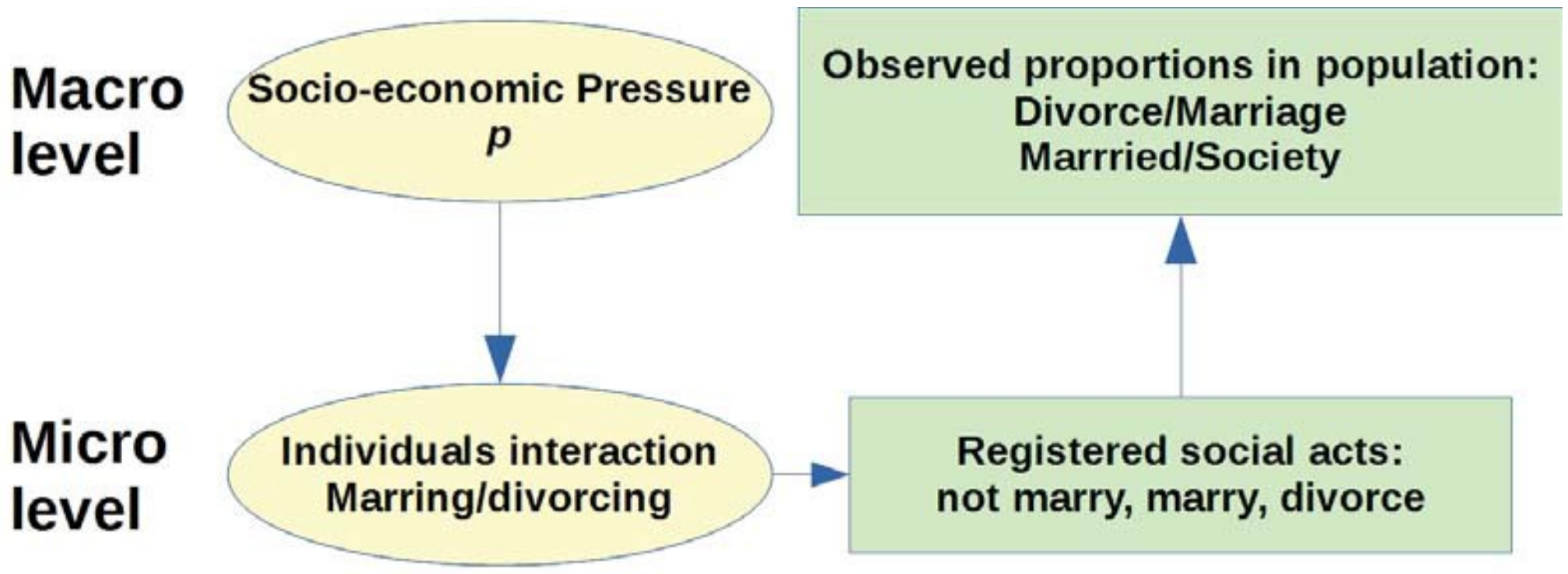

Figure 1: Social action scheme. Socio-economic address individuals preference-based partnership decisions and the outcomes of such decisions come out at the population level.

B) Influence on marriage/divorce

Within Coleman theory of social action [7], we assume existence of macro variable $p$ which drives micro dynamics of individuals decision on marriage or divorce [Fig. 1]. Here, we do not allow our pressure $p$ to evolve in simulation (pressure $p$ trajectory comes from various simulation setups) and do not assume feedback - backward influence on pressure. Individuals marry/divorce for many different reasons from practical to romantic one. However, some external factors could be distinguish [3, 6, 8]:

- Social [3]/economic [8] pressure to have a partner;
- Social/economic pressure to stay in marriage.

Such a global field, we called $p$ - socio-economic pressure, is a combination of social norms and economic benefits of shared household. Environmental context might influence the decision of a individual in many ways. The most important can be grouped by: opportunity-base; belief-based; desire-based and trigger mechanisms [9]. All of these mechanisms describe complexity of human behavior in different aspects. In our paper, we simplify mentioned factors into just one socio-economic parameter pressures of all factors and mechanism due to high level of interactions between them. Moreover, also external impulse could affect socio-economic pressure and important events can rapidly change the pressure. For example, divorce rates increase after natural disaster (releases of social norms), but decrease during economic crises (increase of economic pressure). In middle range theory, "social pressure" also depends on friends influence (as in Ising model of opinion formation). However, the range of that social influence is unknown, so we keep pressure $p$ as a macro variable. Middle range effects of peer was introduced to the model by selection preference. Frequency of matching two single agents was increased in the model, due to homophily in that category [10].

C) Theory of attraction to marry/remarry

People are attracted to partners who are satisfying them. Attraction is estimated on the scale on which subject rates a target person [11]. Each of agents tries to maximize its own satisfaction (find the best partner) with the local knowledge of encounters, he/she meet [12]. We choose 3 dimension by adopting Theory of Capital [13, 14]:

S - social (religious background, political orientation, social status and social skills)

E - economic (property, ability to satisfy the needs, ability to maintain family)

P - physical (physical attractiveness, reproductive ability, health)

In our model, every agent is described in these 3 dimensions. Potential parter is rated subjectively by each person based on its expectation, which could be understood as mix of similarity and complementarity [15]. Moreover, no basic general law works in this field [16], so in our model agent have preferences independent of their own attributes in social (S) and economic (E) dimension. However, there are canons of beauty in every culture in terms of physicality (P) and we decide that agent's preference is equal it's own attribute (homophily).

## 2. Model description

We develop an agent-based marriage model based on population dynamics and social interaction. In the social sciences, the marriage dynamics has been studied widely even with agent-based approach [17]. However, these models require plenty of parameters and results are poorly compared with register-based data [18]. We propose simplified model with one driving macro variable $p$ pressure and few tunable secondary parameters.

In each time step we arrange random encounter between opposite set representatives with priority to pick up single agents. Agents do not know all of their potential partners at the beginning [19], and must meet each others to test match. During each encounter both of the potential partners check their mutual attractiveness [20] and may declare a will to commit to a new partnership (if they have ones, they must first break up their previous relationships). The evolution is described by only one macro parameter $p$ pressure (inverse freedom) of the system representing sexually active part of population of size $N$ ($N/2$ males and $N/2$ females). We distinguish two different executive meaning sof pressure:

- pressure to get into marriage (when individual is forced to engaged with somebody) and we define it as $p$,
- pressure to stay in marriage (when people are forced to stay in marriage) and we define it as $p/scal$, where scaling factor $scal$ was chosen according to statistics of matching distance.

To imitate society as changing creature, we introduce death and birth of agents (there newborns begin life as singles). Every round randomly chosen male agent and female agent are removed with probability $1/aff$ from the system (their relations are also removed). New agents in single state are entering the system in their place, so population size is constant. The average life of agent is $aff$, the average number of encounters. Let assume, that sexually active lifespan is 50 years between [21] age $15-65$ (even it may vary in different societies), and marital activity starts around 5 years later and finish 15 years earlier. An averaged agent has affair opportunity every $30/aff$ year. Some biochemical studies suggest that people can have love affair every 2 years and some psychological studies suggest every 7 years ("The seven-year itch") and such a dynamics was broadly investigated empirically [22, 23], but it's still an open problem for modeling.

We assume also a selective preference within singles. We implemented that by re-sampling the agents for encounter if they both are not singles. We allow up to $l$ re-sampling in every round. Then after $l-th$ re-sampling, we pick up agents apart of their marital state.

### 2.1. Matching rule

We describe attractiveness by vector to imitate its different aspects: physical $P$, social $S$ and economic $E$. Let $P_a, S_a, E_a$ are parameters describing agent's attractiveness, and $P_p, S_p, E_p$ will be parameters to describe agent's preference. Let $P_p = P_a = P$. We choose whole of aspects of attractiveness and preference as unitary distributed random variables in range $[0, 1]$.
The matching between two agents $j, i$ is defined as Manhattan distance:

$$d_{ij} = |S_a(j) - S_p(i)| + |S_p(j) - S_a(i)| + |E_a(j) - E_p(i)| + |E_p(j) - E_a(i)| + |P(j) - P(i)| \quad (1)$$

### 2.2. Dynamics

In every step (round), we randomly select nodes $j$, $i$ (let assume that re-sampling procedure is finished).
A) If both of agents are single and their matching is better then critical value of pressure ($d_{ij} < p$), the new marriage between them is created.
B) If one of chosen agents is married ($i$), and potential new partner ($j$) gives matching better more by critical value of scaled pressure ($d_{ij} < d_{ik} - p/scal$) than it's current partner ($k$), the new marriage between $i, j$ is created and $i$ get divorce with $k$.
If both $i$ and $j$ were married, its possible marriage must be tested for both conditions and ($d_{ij} < d_{ik} - p/scal$ AND $d_{ij} < d_{jm} - p/scal$) must apply to make marriage between $i$ and $j$ (and 2 divorces will take place).
C) In each step randomly selected marriage is tested for survival. Marriage breaks with probability inverse proportional to pressure ($p$). We can understand inverse pressure (freedom) as a thermal noise, which allow agents to spontaneously change their state to less ordered (single). We test this marriage if breaks for:

$$d_{ij} < p + max(0, (1 - p/scal)\,|N(0, 1)|) \quad (2)$$

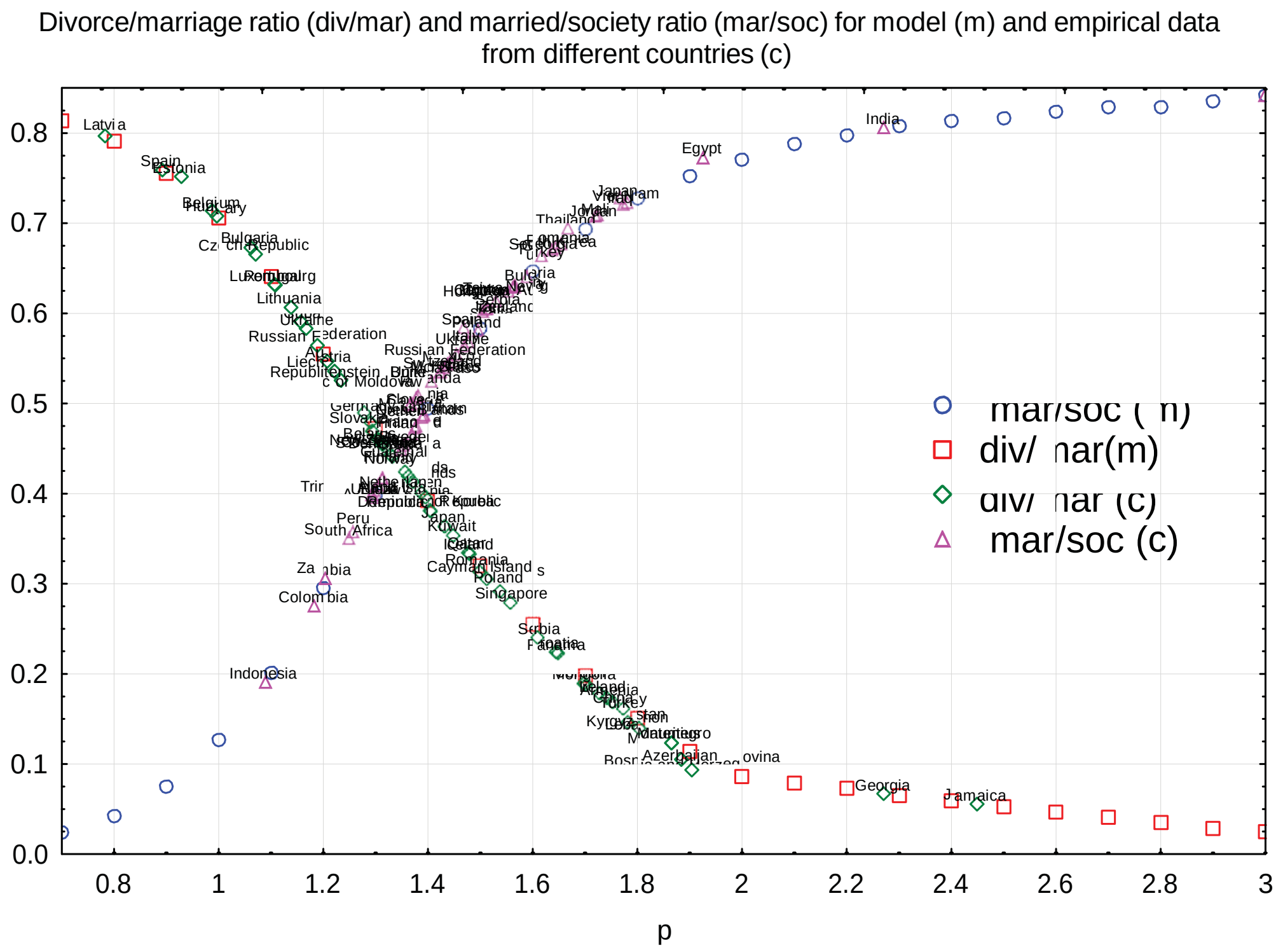

Figure 2: Theoretical pressure values derived from the model for single observable (mar/soc and div/mar) match.

## 3. Results of simulation

Here we present simulations for given choice of control variables: $scal = 3$, $N = 4000$, $aff = 10$, $< l >= 5 * p/3$ (more information about choosing these parameters can be found in Appendix). Computer asynchronous simulation investigations focus on the statistical properties of the outcomes, which describe real world societies like: proportion of divorces to marriages, proportion of married in the society.

### 3.1. Stationary case

We check the properties of our system when its goes to equilibrium. For different range of $p$ it takes different time (for small $p$ shorter, but for large longer). We adjust pressure parameter to register-based data from various

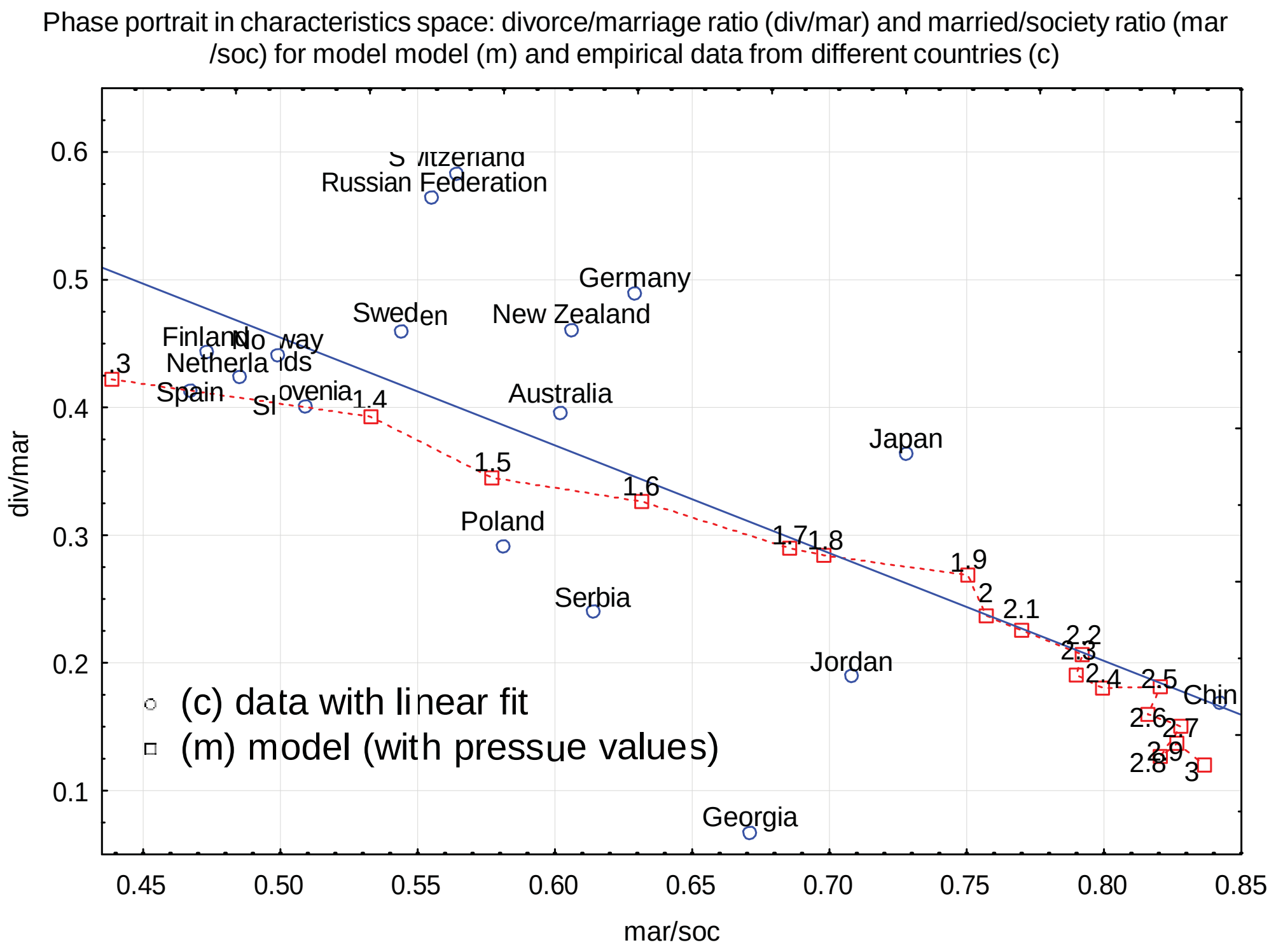

Figure 3: Theoretical pressure in model and countries observables on phase plane

countries [Fig. 2] separately for two observables. We use empirical values as: proportion of married in the society *mar/soc* from WVS (Wold Value Survey) in the latest available call (2012 or 2007) and proportion of divorces per marriages *div/mar* from UN (United Nations) registry in also the latest available edition ($2007 - 2010$). First nontrivial observation characterize system behavior for very small pressure (with some evolutionary population reproducing issues). If there is no pressure ($p = 0$) agents prefer single state (almost 100% of population) and if they marry, the marriages do not survive for too long. Too small values of $p$ seem to be non-physical [Fig. 2] (first countries are places from $p > 0.75$). For high values for $p > 3$ ratio of married in population (*mar/soc*) is saturating on the natural limit (around 0.9) and cannot increase due to death and birth mechanisms. We observe the phase transition in preferential state (*mar/soc*) with increase of pressure from single to coupled. There is also phase transition in proportion of divorces to marriages (*div/mar*), it decease in all range of $p$. *Div/mar* is close to 1 for $p \sim 0$ and for $p$ going to infinity it suppose to be 0. The empirical range of our observables: $0.05 < div/mar < 0.8$ and $0.2 < mar/soc < 0.85$

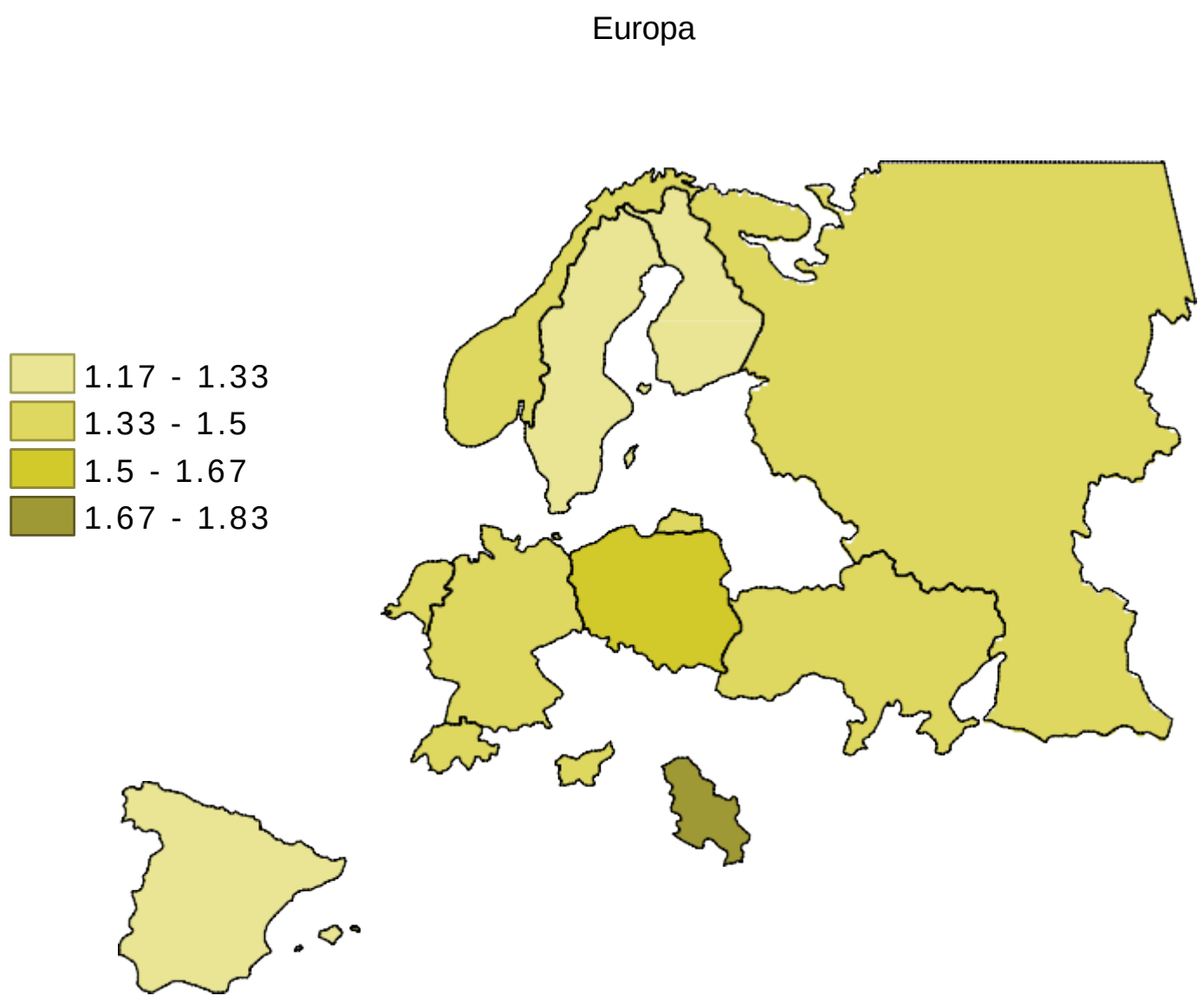

Figure 4: Pressure estimates in available European countries

### 3.2. Pressure estimates

We try to estimate pressure in countries based on comparison of real-life statistics with model outcomes. For final classification, we include countries for which statistics from each all divorce, marriage and population structural values were available for year 2010 in United Nations (UN) database and in Wold Value Survey (WVS) call from 2012. Theoretical pressure line divides phase space into two regimes upper and lower bound [Fig. 3]. Most of the countries are placed close to this line, which somehow confirm basic agreement of our model with reality. In upper bound ($\frac{mar/soc}{div/mar}$ is low enough), there are countries with under-representatives of married (other kind of legal partnerships like in Sweden; high number of widows like in Russia, etc). In lower bound divorces are under-represented (law restrictions like in Georgia). However, pressure combine allocation is less sensitive to country-specific idiosyncratic situation. In the two dimensional space (*mar/soc*, *div/mar*) [Fig. 3], we calculate the shortest Euclidean distance to theoretical trajectory *p* (in terms of least squares) for each country. We present European countries with our 2-dimensional pressure estimates on the map [Fig. 4]. Estimates are also in agreement with demographic transition theory [5], because conservative Post-communistic countries as Poland or Serbia (during second demographic transition) are characterized by high pressure and liberal Nordic countries or Spain (after second demographic transition) by low *p*.

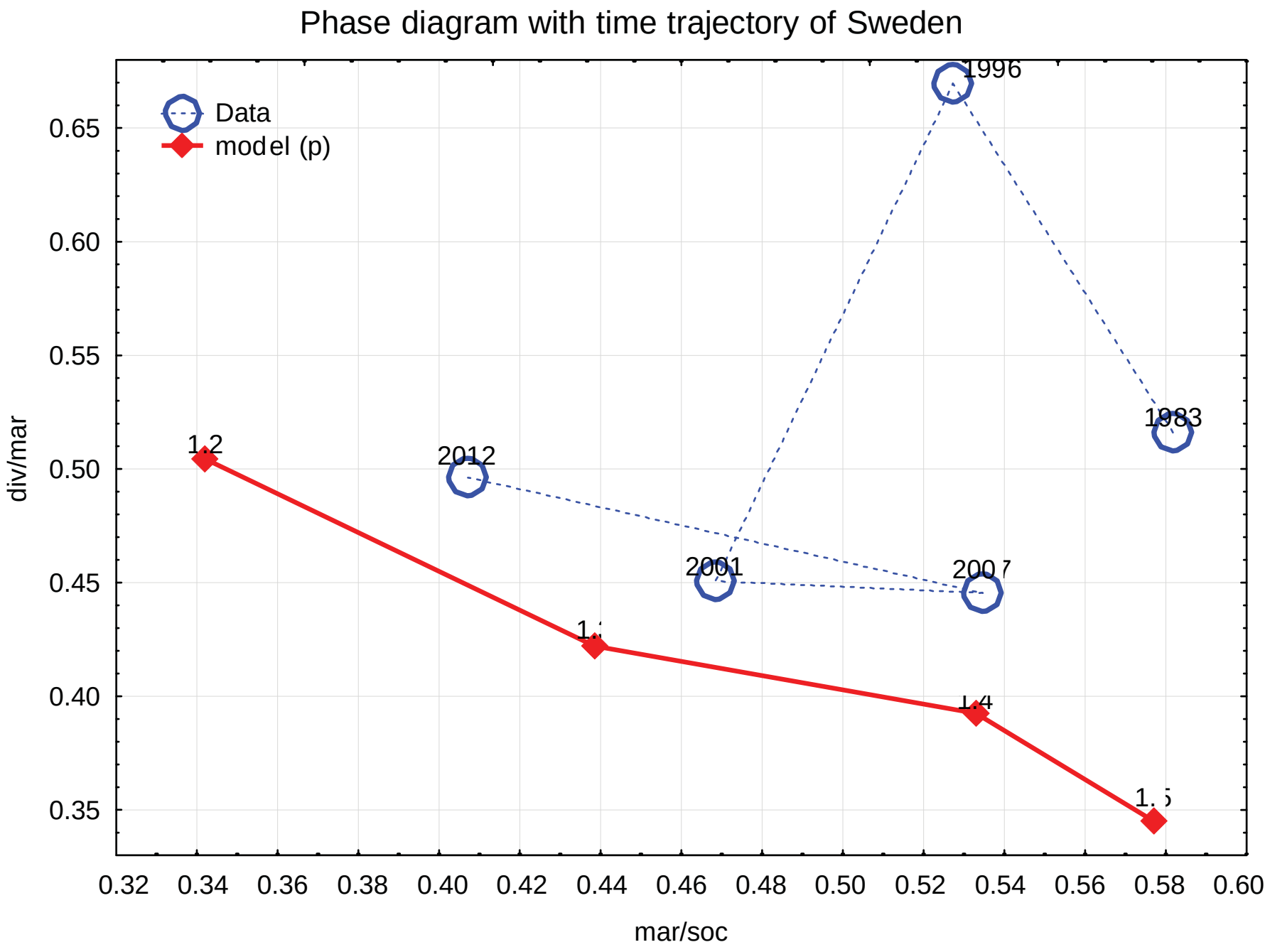

Figure 5: Phase diagram with time evolution

### 3.3. Time evolution portrait

We observe time evolution of Swedish population statistics. Our simplified model does not allow to explicit time analysis and *p* is constant. However, we can separately estimate pressure parameter for a given time point and observe how it changes with time. On phase diagram we can trace the change of direction in Sweden [Fig. 5] and overall trend leads into decreased pressure in last 30 years. However, process triggers around Millennium and continue into less pressured regime recently. Moreover, divorce rates are lower now than 30 years ago. This effect cannot be explained non from social action as well as demographic theory, so we discuss non-homogeneous society hypothesis later in conclusion.

## 4. Conclusion, Discussion and future works

In previous studies, only divorces or only marriages were investigated and its first (up to our knowledge) agent-based study for both aspects of global societies. We provide evidence that cross-country variation in socio-economic pressure $p$ can explain the observed variation in divorce/marriage and married/society proportions in some extend. We claim, that our simple model reflects norm diversification in the world, but many problematic aspects should be arise. Firstly, we assume homogeneity of society and its already known to false. From homogeneity assumption we observe exponential or Gaussian distribution of agent properties (e.g. number of affairs or marriages), but in real world other distributions as long tailed (log-normal or power law) dominate in that aspect. In this paper, we analyze only mean values due to that bias. Observed real-values like proportion of divorce to marriage and faction of married in a society are sensitive to demographical changes and age-structure of investigated population.

In further works, dynamical aspects should be also deeply investigate in order to test some hypothesis on First and Second Demographical Transition as well as parallel observed processes of liberalization and radicalization. Moreover, some parallel, significantly different traits are circulating in many Western societies due to their multicultural structure. Our single pressure parameter additionally corresponds to both social and economic pressure. For example in Sweden, immigrants with a non-Western background were found to be less prone to local social norms [24], but on the other hand they perfectly suit to pro-family economic benefits. Moreover quasi-chaotic trajectory around Millennium in Sweden [Fig. 5] could be an effect of several waves of conservative immigrants and future pressure prediction in not trivial, because of the inverse liberalization.

Another issue is strict division between heterosexual, monogamist married and singles due to simplicity and accessibility of data. In real world other kind of relation coexist, like homosexual or cohabitation, but in this methodology their would be classified into singles. In fact, for some countries more detailed data are available, and additional states seem to have very interesting implications.
We would like also discuss meaning of 'unphysical' regime for low pressure value [Fig. 2, 3]. If agents do not want to tie, the society is likely to die out if it would not introduce replacement institution to a family, which was historical fundamental of reproduction. Long-term marriage gives evolutionary, the best fertility rates and even if its not a ground true any more (there are some exceptions, e g. in a theoretical postmodern society). Unfortunately, in this model non-marital forms of relations, which also could play reproductive functions, are not include. Currently, we observe drift of pressure parameter to this forbidden region. The country closest to that region [Fig. 2] is Latvia (single observation estimate of pressure), which has one of the lowest fertility in the world. We speculate, that if reproductivity outside of traditional family institution would be not possible, some societies below critical value of pressure would extinct.

## Appendix A. APPENDIX

The model description provided in this paper is probably the best choice from a set of candidate models and whole spectrum of parameters. Here, we provide additional information: in case of matching two single agents, divorce scaling and alternative formulation.

### *Appendix A.1. Matching distances $d_{ij}$ and its remarks on control variables p*

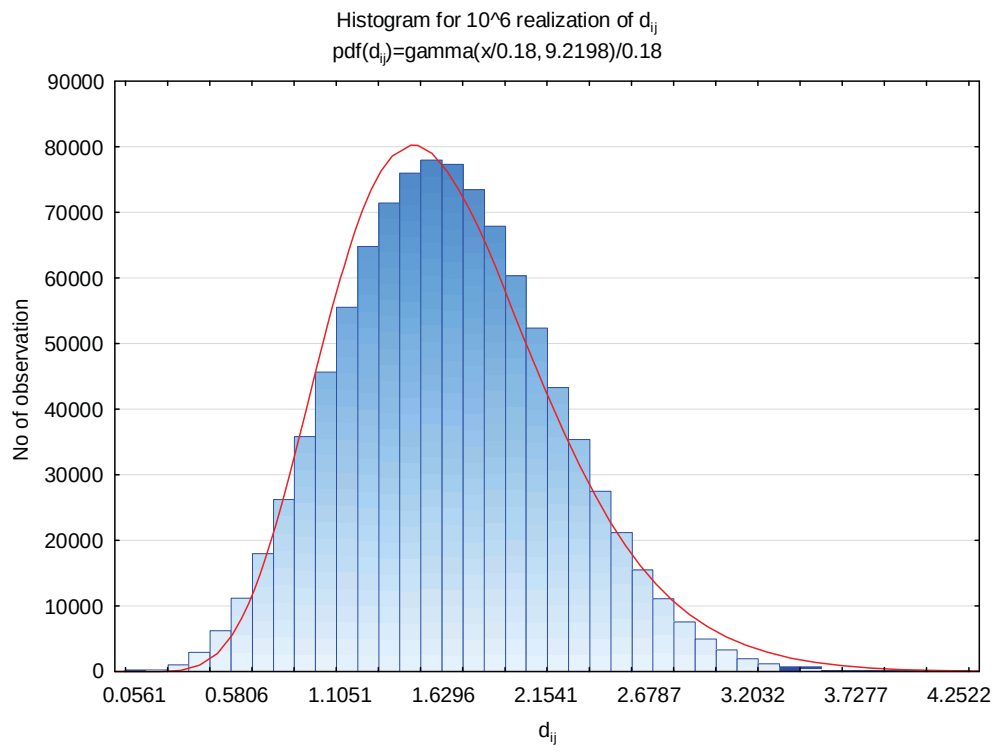

Figure A.6: Statistics of matching distance $d_{ij}$

Provided in the main text definition of $d_{ij}$ limits range of possible matching [Fig. A.6]. There are some natural thresholds for matching distances $d_{ij}$. Minimal factor $p = 0.5$ was chosen as a minimal value for each agents can decide to marry (0.5 percentile). Otherwise almost none of affair would finish with marriage. On the other hand, for $p$ greater than critical value $p = 3.08$ (99.5 percentile), every affair of two single agent will come up with marriage.

### Appendix A.2. Estimating scal and type of spontaneous breakage

The scaling factor $scal$ stands for pressure to stay with the same partner and represents the subjective strength of marriage. We assume that being married decrease attraction to a new partners. For $scal < 1$ divorces would be statistically impossible and if $scal > 6.3$ only 5% of marriage in danger (when a potential new parter is at least as good as previous one) will survive. The range of our interest should be in the middle of presented extremes. We look for few scenarios - namely for $scal = 2$, 2.5 or 3. For small $scal$ and $p$ large enough divorces stop at some point when agents are more less satisfied with their partners and cannot change them anymore.
Additionally, we tested spontaneous breakage scenario presented in the main text as well as alternative one with lower potency. In this alternative scenario ('alter noise') we test in every round two agents (female and male):

$$d_{ij} < p_f + max(0, (1 - p/3)N(0, 1))/scal \qquad (A.1)$$

We find the best fit to data for $scal = 2$ and with standard spontaneous breakage [Fig. A.7]. In the alternative scenario, thermal effect appears with a little extend only. Spontaneous breakage in the original form increases visually the fit, mainly for low $p$. This can be interpreted as follows: it is easier to change agent state at higher temperatures (e.g. more noise ~ higher temperature), so pressure trajectory should be steep for small $p$.

### Appendix A.3. Estimating aff and l

The interaction between $aff$ and $l$ has been investigated due to the optimal choice of these variables. We search within suggested by empirical studies number of affairs between 5 to 20 livelong. In asymptotic case for pressure large enough to tide every pair in each affair, and then divorce are not possible at all, formula for ratio of married in population ($x = mar/soc$) could be derived from flow equilibrium. We are looking for a stable point where proportion of married in population is almost constant ($dx/dt == 0$).

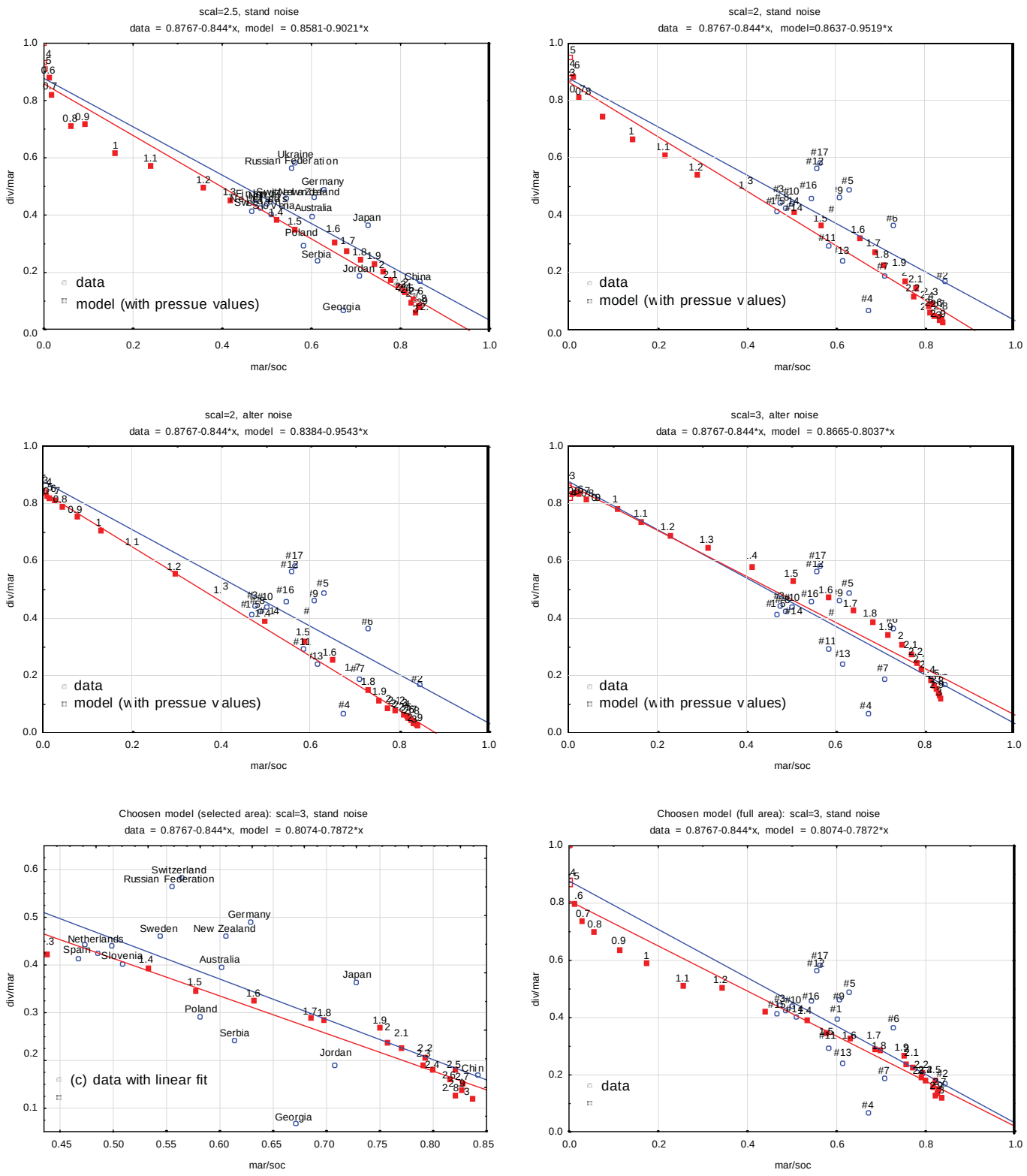

Figure A.7: Sensitivity of scaling parameter ***scal*** (values 2, 2.5 and 3) and choice of noise (alternative or standard). Linear fit for data and model is also provided

Let us extend: $x = M/(M+S)$ and $N = M+S$, where: M - number of married and S- number of singles. In one cycle of simulation $W_{M-S} = Nx/aff$ agents move from M into S due to dead/birth process. The same time $W_{S-M}$ agents get married. To do so he/she must has an affair with other single agents and the probability that indicated single agent meet another single is equivalent to sampling combinations with repetition. It comes from the rule, that singles attract single, so we redraw up to $l$ times if our encounters and not both singles.

$P_{S-M} = 1 - \left(\frac{C^l_{M/2}}{C^l_{N/2}}\right)$, where $C^a_b$ is a number of combinations with repetitions size $a$ from set of elements $b$. We simultaneous (independently) draw pair of heterosexual agents, so this probability is squared. After comparing $W_{S-M} = W_{M-S}$ and simplifying equation we get:

$$x/aff = \left(1 - \frac{\prod_{i=1}^{S}(M/2+i-1)}{\prod_{i=1}^{S}(l+M/2+i-1)}\right)^2 \qquad (A.2)$$

For given parameters space, we indicate regions [Fig. A.8], where $x$ reflects possible real-life values for very high pressure. Thus, $x > 0.85$ (the highest empirical observation in WVS from China). This condition is always satisfy for ***aff*** $>$ 10, so we choose ***aff*** $= 10$. The best choice of $l$ lays around 5. While redrawing is do not need within low pressure systems, we decided to define $l = 5p/3$, which for critical $p \sim 3$ gives us $l \sim 5$.

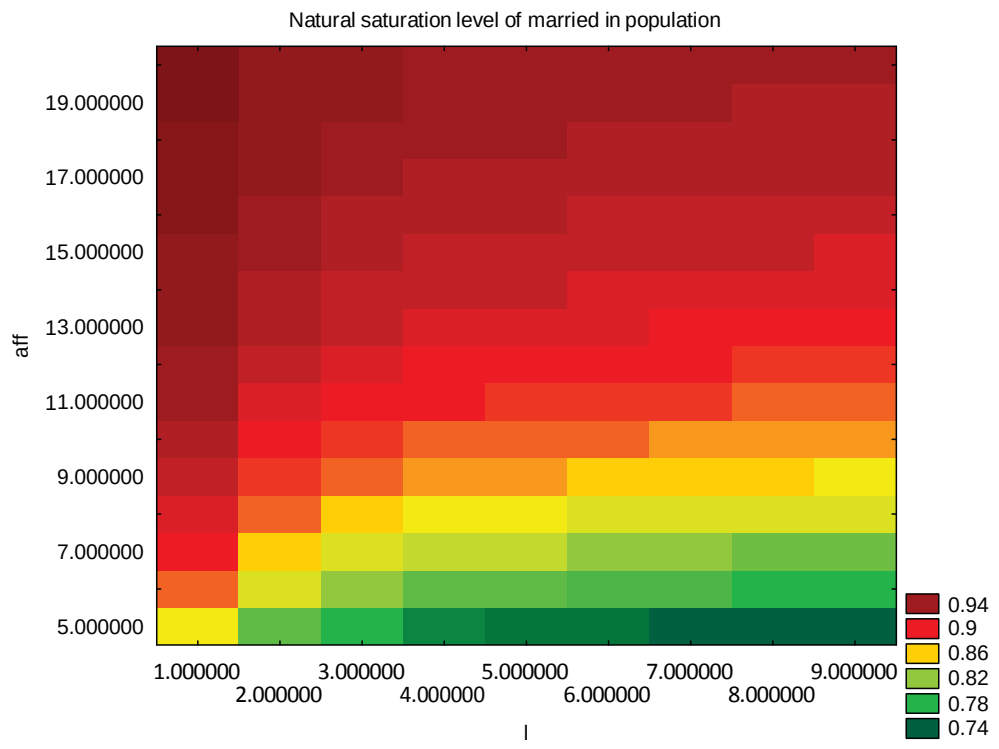

Figure A.8: Saturation level of married in population. Real life saturation level reaches 0.9

### *Appendix A.4. Relaxation times and size scaling*

We did not find criticality in phase transition between coupled and single phases in size scaling. Regression coefficients (tangents of angle in transition region) seems to do not differ respectively to system size [Fig. A.9].

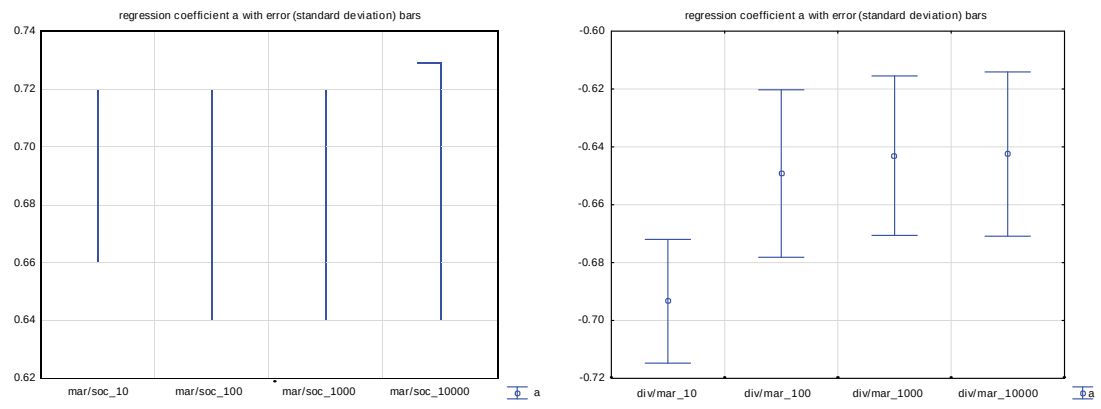

Figure A.9: Lack of system size dependence. Speed of phase transition in both dimensions: ***mar/soc*** and ***div/mar*** for different system sizes.